\documentclass[twocolumn,showpacs,pra,aps]{revtex4}

\usepackage{epsfig}
\usepackage{graphicx}
\usepackage{dcolumn}
\usepackage{bm} 
\newcommand{\beq}{\begin{equation}}
\newcommand{\eeq}{\end{equation}}
\newcommand{\beqa}{\begin{eqnarray}}
\newcommand{\eeqa}{\end{eqnarray}}

\newcommand{\om}{\omega}

\def\jpb#1{{ J.\ Phys.\ B} {\bf#1}}

\def\natphys#1{{ Nature\ Phys.\ } {\bf#1}}
\def\njp#1{{ New J.\ Phys.\ } {\bf#1}}
\def\pra#1{{ Phys.\ Rev. A\/} {\bf#1}}

\def\prl#1{{ Phys.\ Rev.\ Lett.} {\bf#1}}

\begin{document}

\title{Classical Theory of High Field Atomic Ionization Using Elliptical Polarization}

\author{Xu Wang}
\author{J.\ H.\ Eberly}
\affiliation{ Rochester Theory Center and the Department of Physics
\& Astronomy\\
University of Rochester, Rochester, New York 14627}

\date{\today}

\begin{abstract} Important information about strong-field atomic or molecular ionization can be missed when using linearly polarized laser fields. The field strength at which an electron was ionized, or the time during a pulse of the ionization event are examples of such missing information. In treating single, double, and triple ionization events we show that information of this kind is made readily available by use of elliptical polarization.
\end{abstract}

\pacs{32.80.Rm, 32.60.+i}

\maketitle


\section{Introduction}

The interaction between intense laser fields and gas phase atoms or molecules has attracted attention for the past two decades \cite{Becker-Rottke}. Many strong-field phenomena have been observed and controlled, such as ionization of atoms \cite{Walker-etal}, dissociation of molecules \cite{Bandrauk89}, generation of high harmonics \cite{Lewenstein-etal} and creation of attosecond pulses \cite{Krausz-Ivanov}.

To make a theoretical description of the interaction between an intense laser field and a multiple-electron atom is not an easy task. First, no analytical quantum mechanical solutions can be expected. Second, the laser electric field strength is comparable to the atomic Coulomb electric field strength felt by a valence electron, so neither the laser field nor the Coulomb field can be regarded as a small perturbation and familiar perturbation theories cannot be used. Third, full-dimensional numerical calculation of the time-dependent Schr\"odinger equation (TDSE) is extremely demanding in computational resources and is effectively limited to the helium atom \cite{Parker-etal96,Parker-etal06}.

Simplified theories and models are thus desirable and have been developed to match rapidly emerging experimental results (for a review of such theories, see \cite{Becker-etalRMP}). Among these theories, a semiclassical three-step recollision model has been widely used to understand various strong-field phenomena heuristically \cite{Corkum, Kulander}. However, interesting questions (e.g., the proper understanding of electron release times \cite{Pfeiffer-etalNatPh}) are being raised in high-field atomic photoionization that have little or nothing to do with recollision. Here we report theoretical calculations concerning single, double and triple ionization for laser intensities in the PW/cm$^2$ range with elliptically polarized pulses and without recollision.

\begin{figure}
  \includegraphics[width=4cm]{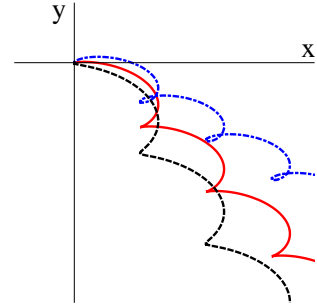}\\
  \caption{An illustration of Simpleman electron trajectories in an elliptically polarized laser field with ellipticity 0.5. Three close trajectories are shown, corresponding to three slightly different initial velocities $v_y$ at the time of emission.}\label{f.Simpleman}
\end{figure}

Two theoretical approaches have been used. An analytical theory is first developed for the sake of physical clarity by extending the Simpleman theory \cite{Simpleman} to include elliptical polarization. An illustration of Simpleman electron trajectories in an elliptically polarized laser field is shown in Fig. \ref{f.Simpleman}. Numerical calculations are then performed using a classical ensemble method \cite{Panfili-etal}, which treats the entire system, both the laser field and the atom, purely classically but going beyond the Simpleman approach. We do this by including all forces (electron-ion and electron-laser) in obtaining solutions of the time dependent Newton equations (TDNE).

The classical ensemble method has provided valuable insights into ionization dynamics \cite{Ho-etalPRL05,Ho-Eberly06,Ho-Eberly07,Haan-etal08,MCU09,MCU10}. It takes into account naturally the occurrence or the non-occurrence of recollision, so it can be used to study strong-field questions with or without recollision. Recently it has been extended to include elliptical polarization and physical processes without recollision \cite{Wang-EberlyPRL09} and good agreement with experiment \cite{Maharjan-etal} has been achieved.

The purpose of this paper is to show that elliptical polarization has the ability to uncover ionization information that is otherwise unreachable with linear polarization. For example, under what field strength was an electron ionized? At what time during the pulse was an electron ionized? Note that an electron may not be ionized at the pulse peak if the peak intensity of the pulse is higher than the intensity needed to ionize this electron. This is especially notable for intense laser fields that are capable of ionizing more than one electron. This ionization information cannot be directly and easily obtained with linear polarization. However, as we show, this kind of information can be straightforwardly obtained from the end-of-pulse ion momentum distributions obtained under elliptical polarization. Ion momentum distributions can be measured using the COLTRIMS (cold target recoil ion momentum spectroscopy) technique \cite{COLTRIMS}. We will show that the momentum distribution of a singly charged ion reveals the ionization field and the ionization time of the emitted electron. The momentum distribution of a doubly charged ion reveals the ionization fields and the ionization times of both emitted electrons. And the momentum distribution of a triply charged ion tells the ionization fields and the ionization times of all three emitted electrons.

Such ionization information may reveal new ionization dynamics. For example, will electrons really follow the ionization fields and the ionization times predicted by independent-electron tunneling formulas \cite{ADK,Tong-Lin}? A recent experiment performed on argon has given a preliminary answer no, although the detailed physics is still under investigation \cite{Pfeiffer-etalNatPh}.

The following questions can be answered by future experiments using elliptical polarization: How good are the commonly used tunneling formulas? What is exactly the role of the remaining electrons during an ionization process? Does this role depend on atomic or molecular species? Is this role the same for the first ionization process and for the second ionization process, and even for the third ionization process?

This paper will be organized as follows. In section II we will first review the Simpleman analytical theory to explain how one can use elliptical polarization to obtain the above-mentioned ionization information. In section III we go beyond Simpleman theory and use the classical ensemble method to perform numerical TDNE experiments and to test the accuracy of the analytical theory for an atom with three active electrons. In section IV results of the numerical experiments will be shown and compared to the predictions of the analytical theory. Summaries are presented in section V.

\section{Simpleman theory}

In this section we recall a simple analytical theory that links the experimental ion momentum distributions with the ionization information interested. This so-called ``Simpleman" theory \cite{Simpleman} has long been used to understand strong-field ionization processes, especially electron kinematics after emission from the parent ion.

\begin{figure*}[t!]
  \includegraphics[width=5cm]{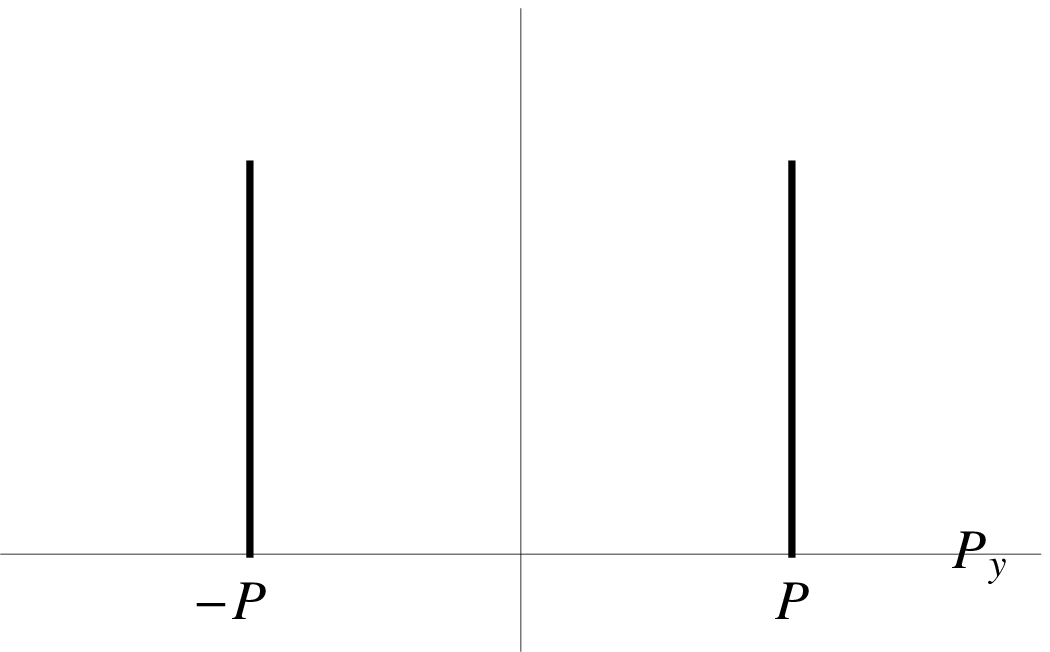}
  \hspace{0.5cm}
  \includegraphics[width=5cm]{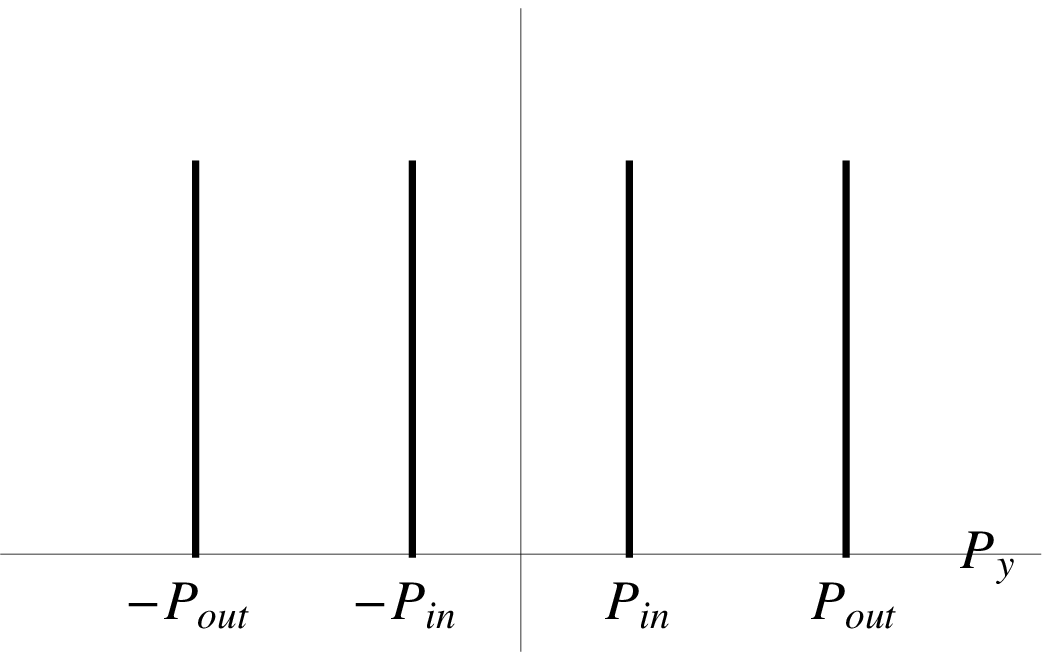}
  \hspace{0.5cm}
  \includegraphics[width=5cm]{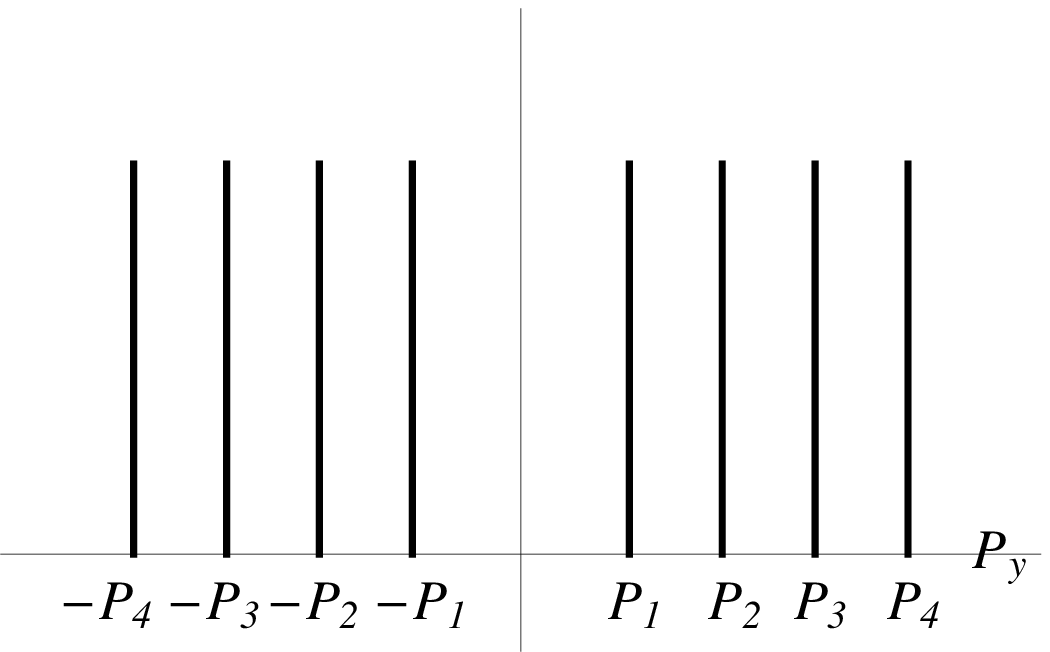}
  \caption{Schematic illustration of the ion momentum distribution along the y direction for single ionization (left), double ionization (center), and triple ionization (right). This figure is only for the purpose of demonstration, so it is not drawn to any scale and no further complications like the peak widths have been taken into account.}\label{f.248peaks}
\end{figure*}

The Simpleman theory starts from the time that an electron is just emitted and neglects the ion core Coulomb potential. The ionized electron is treated as a classical particle and its motion is governed by classical mechanics, via solutions of the TDNE
\beq
    \vec{F} = m \vec{a} \quad \text{and} \quad \vec{F} = q \vec{E}_L (t),
\eeq
where $\vec{E}_L(t)$ is the laser electric field.

The momentum of the electron at the time of ionization is approximated to be zero: $p(t_1) = 0$. The momentum of the electron at some later time $\tau$ (for example, at the end of a pulse) is just
\beq
    \vec{p} (\tau) = \int_{t_1}^{\tau} \vec{a}(t) \mathrm{d} t = - \int_{t_1}^{\tau} \vec{E}_L(t) \mathrm{d} t. \label{e.ptau}
\eeq
Atomic units are used. The charge of the electron is -1 a.u. Note that for the intensities that we are interested in, relativistic effects are negligible and the magnetic part of the Lorentz force can be ignored.

The Simpleman theory has mostly been used for single ionization with linear polarization. Here we will extend it to include elliptical polarization and to take into account single, double, and triple ionization. Let us start from the ionization of a single electron in an elliptically polarized laser field
\beq
    \vec{E}_L(t) = E_0 f(t) \left[ \hat{x} \sin(\omega t + \phi) + \hat{y} \varepsilon \cos(\omega t + \phi) \right], \\ \label{e.field}
\eeq
with
\beqa
    E_x(t) &=& E_0 f(t)\sin(\omega t + \phi), \\
    E_y(t) &=& \varepsilon E_0 f(t) \cos(\omega t + \phi).
\eeqa
Where $f(t)$ is the pulse envelope function, $\omega$ the angular frequency, $\phi$ the carrier envelope phase (CEP), and $\varepsilon$ the field ellipticity. We choose the x direction as the major polarization direction and the y direction as the minor polarization direction (recall Fig. \ref{f.Simpleman}).

Suppose an electron is ionized at time $t_1$ with zero velocity, and suppose that the ion core Coulomb attraction can be neglected after ionization. Then the momentum of this electron {\it at the end of the pulse} can be straightforwardly derived
\beqa
    p_{1x} &=& -\frac{1}{\varepsilon \omega} E_y (t_1) \approx 0; \label{e.SIPx} \\
    p_{1y} &=& \frac{\varepsilon}{\omega} E_x (t_1) = \pm \frac{\varepsilon}{\omega} E_1. \label{e.SIPy}
\eeqa

An adiabatic condition has been applied. The duration of the pulse is assumed to be much longer than one optical cycle. Then the end-of-pulse momentum of the electron does not depend on the detailed pulse shape. We see an interesting crossing relation: the end-of-pulse momentum along the x direction ($p_{1x}$) is determined by the instantaneous laser field strength along the y direction at the time of ionization, $E_y (t_1)$, and the end-of-pulse momentum along the y direction ($p_{1y}$) is determined by the instantaneous laser field strength along the x direction at the time of ionization, $E_x (t_1)$.

\begin{figure} [b!]
  \includegraphics[width=5cm]{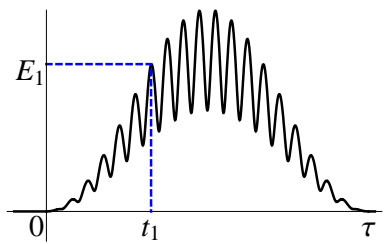}\\
  \caption{Illustration determining $t_1$ after having obtained $E_1$. A sine square pulse with ellipticity value 0.5 is shown. The position of the first field peak corresponding to $E_1$ is used as an estimate of $t_1$.}\label{f.E2T}
\end{figure}

Note that for elliptical polarization, the peak field strength along the x direction is stronger than that along the y direction. Due to the fact that ionization probability depends exponentially on field strength \cite{ADK}, ionization happens most probably around field maxima in the x direction. At such times, the field along the y direction is near zero. Therefore $E_x (t_1)$ can be regarded as the instantaneous laser field at the time of ionization. Since $E_x (t_1)$ could be positive or negative, a ``$\pm$'' sign has been added in front of $E_1$, which denotes the absolute laser field strength at the time of ionization and is always positive.

We can see that the end-of-pulse momentum distribution of the electron, or equivalently of its parent ion, should have a single peak structure centered at zero along the x direction and a double peak structure centered at $\pm \varepsilon E_1 / \omega$ along the y direction. Thus we have obtained a clear relation between the peak positions of ion momentum distribution along the y direction and the ionization field of the electron.

If the ion momentum distribution is measured experimentally using COLTRIMS, projection of ion momentum onto the y direction should give two symmetric peaks, as illustrated in the left panel of Fig. \ref{f.248peaks}. From the positions of the two peaks, noted as $\pm P$ with $P>0$, one gets the ionization field of the electron as
\beq
    E_1 = \frac{\omega}{\varepsilon} P. \label{e.E1Peak}
\eeq
This formula demonstrates why elliptical polarization has the ability to uncover ionization information unreachable with linear polarization: elliptical polarization provides an additional dimension, which contains information.

The corresponding ionization time can be inferred assuming a smooth pulse shape, e.g., a gaussian or a sine-squared pulse envelope. Figure \ref{f.E2T} illustrates the idea using a sine-squared pulse envelope and ellipticity value 0.5, which is the value used for illustration throughout this paper. Note that in general, there is no one-to-one correspondence between $E_1$ and $t_1$. The method that we have adopted to get $t_1$ in this paper is to use the position of the first field peak corresponding to $E_1$, as illustrated in Fig. \ref{f.E2T}. CEP phase will be averaged out.

Next, let us move one step further for double ionization. Double ionization can be roughly divided into two categories, namely, sequential double ionization (SDI) and non-sequential double ionization (NSDI). SDI implies that the two electrons are ionized one by one without noticeable mutual correlations. NSDI means that the two electrons are ionized almost at the same time with substantial mutual correlations. Recollision is generally conjectured as the physical mechanism that induces the mutual electron correlations present in NSDI \cite{Corkum}. For the ellipticity value used in this paper, the field along the y direction drives the emitted electrons transversely and effectively eliminates the possibility of recollision, so all double ionization obtained can be regarded as originating from sequential processes. Therefore the above argument for the first ionized electron also applies to the second ionized electron. The momentum of a resultant doubly charged ion equals the sum of the momenta of the two ionized electrons
\beqa
    P_x &=& p_{1x} + p_{2x} \nonumber \\
        &=& -\frac{1}{\varepsilon \omega} ( E_y (t_1) + E_y (t_2) ) \approx 0; \label{e.DIPx} \\
    P_y &=& p_{1y} + p_{2y} \nonumber \\
        &=& \frac{\varepsilon}{\omega} ( E_x (t_1) + E_x (t_2) ) = \frac{\varepsilon}{\omega} (\pm E_1 \pm E_2). \label{e.DIPy}
\eeqa

The ion momentum distribution would also be expected to have a single peak structure centered at zero along the x direction. What is interesting is the momentum distribution along the y direction. As Eq.(\ref{e.DIPy}) shows, the ion momentum distribution along the y direction is expected to have four peaks positioned at $\pm \varepsilon (E_1+E_2)/\omega$ (two outer peaks) and at $\pm \varepsilon (-E_1+E_2)/\omega$ (two inner peaks) \cite{Maharjan-etal,Wang-EberlyPRL09}. As in the case of single ionization, we also get a clear relation between the ion momentum distribution and the ionization fields of the emitted electrons.

If the doubly charged ion momentum distribution is measured experimentally using COLTRIMS, projection onto the y direction should give four peaks, as demonstrated in the middle panel of Fig. \ref{f.248peaks}. From the positions of the four peaks, noted as $\pm P_{out}$ and $\pm P_{in}$ with $P_{out}>0$ and $P_{in}>0$, one gets the ionization fields of both emitted electrons as
\beqa
    E_1 &=& \frac{\omega}{2\varepsilon} (P_{out} - P_{in}); \label{e.DIE1} \\
    E_2 &=& \frac{\omega}{2\varepsilon} (P_{out} + P_{in}). \label{e.DIE2}
\eeqa
The ionization times of the two electrons can also be obtained numerically using a similar method as illustrated in Fig. \ref{f.E2T}.

The same strategy can also be applied to triple ionization. The ionization fields and the ionization times of all three ionized electrons can be obtained from the experimentally measured ion momentum distribution. Ideally, one would expect eight peaks along the y direction, as demonstrated in the right panel of Fig. \ref{f.248peaks}. The positions of the eight peaks are labeled as $\pm P_1$, $\pm P_2$, $\pm P_3$, and $\pm P_4$, where
\beqa
   P_1 &=& \frac{\varepsilon}{\om} \left( - E_1 - E_2 + E_3 \right); \\
   P_2 &=& \frac{\varepsilon}{\om} \left( + E_1 - E_2 + E_3 \right); \\
   P_3 &=& \frac{\varepsilon}{\om} \left( - E_1 + E_2 + E_3 \right); \\
   P_4 &=& \frac{\varepsilon}{\om} \left( + E_1 + E_2 + E_3 \right).
\eeqa

There are four equations with only three unknown variables, so knowing any three of $\{P_1, P_2, P_3, P_4\}$ the fourth one can be calculated. The ionization fields $E_1$, $E_2$ and $E_3$ can be deduced from the positions of the peaks
\beqa
   E_1 &=& \frac{\om}{2\varepsilon} (P_4 - P_3) = \frac{\om}{2\varepsilon} (P_2 - P_1); \\
   E_2 &=& \frac{\om}{2\varepsilon} (P_4 - P_2) = \frac{\om}{2\varepsilon} (P_3 - P_1); \\
   E_3 &=& \frac{\om}{4\varepsilon} (P_1 + P_2 + P_3 + P_4).
\eeqa

\section{numerical experiments}

\begin{figure} [b!]
  \includegraphics[width=5cm]{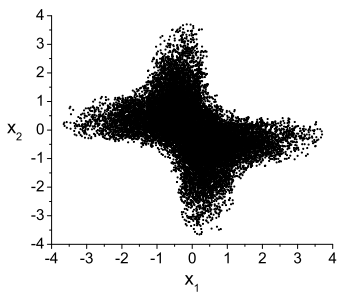}
  \caption{Projection of a classical ensemble onto the x$_1$-x$_2$ plane. Each black dot represents a model atom. The butterfly shape is a manifestation of mutual electron repulsion.}\label{f.ensemble}
\end{figure}

\begin{figure*}
  \includegraphics[width=4cm]{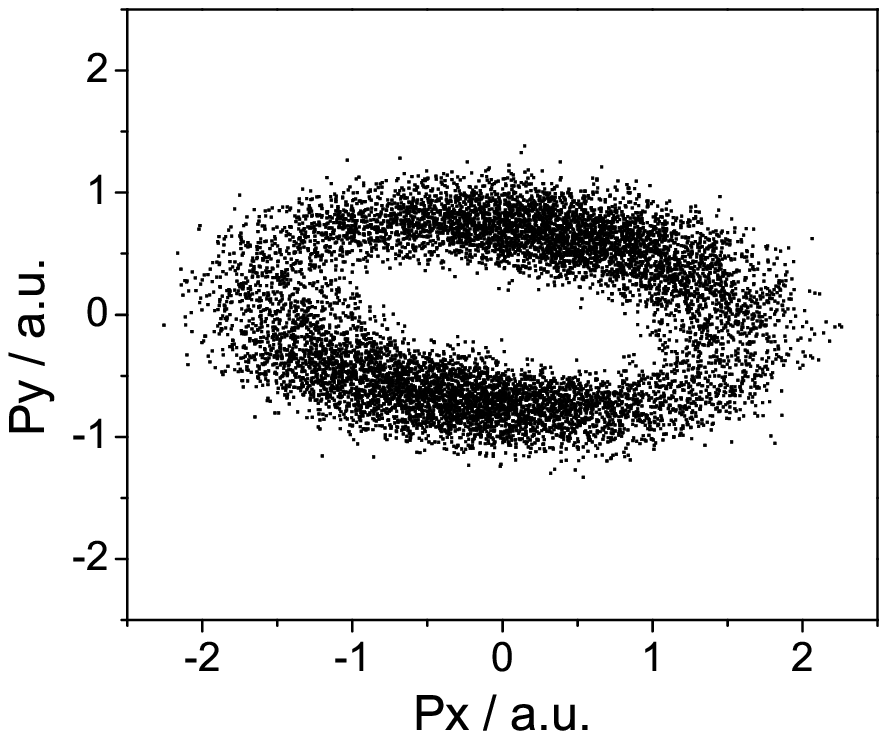}
  \includegraphics[width=4cm]{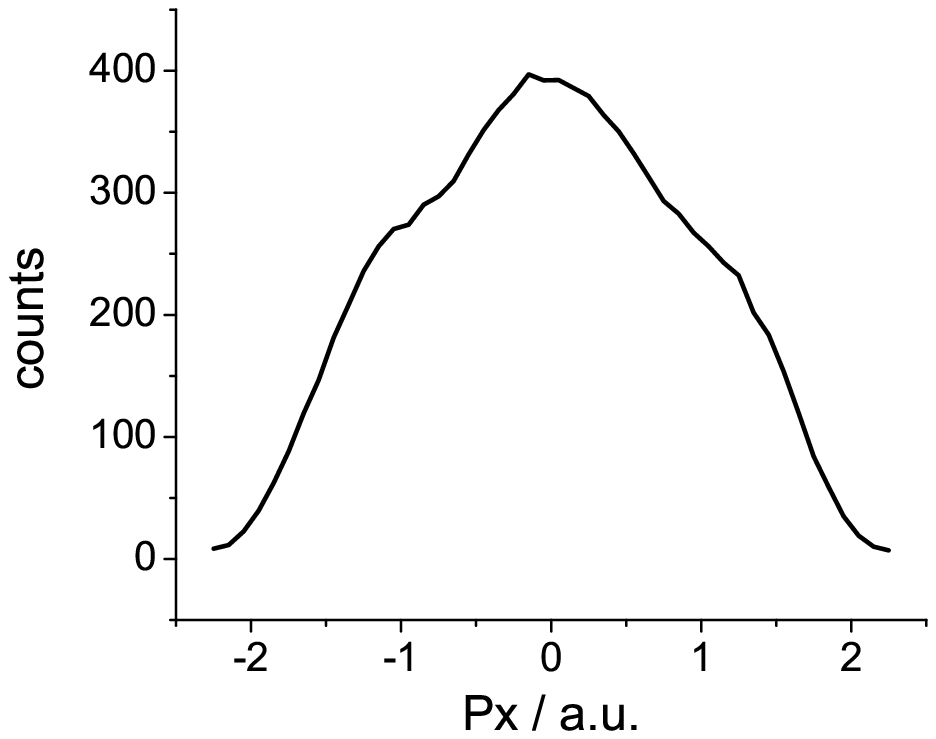}
  \includegraphics[width=4cm]{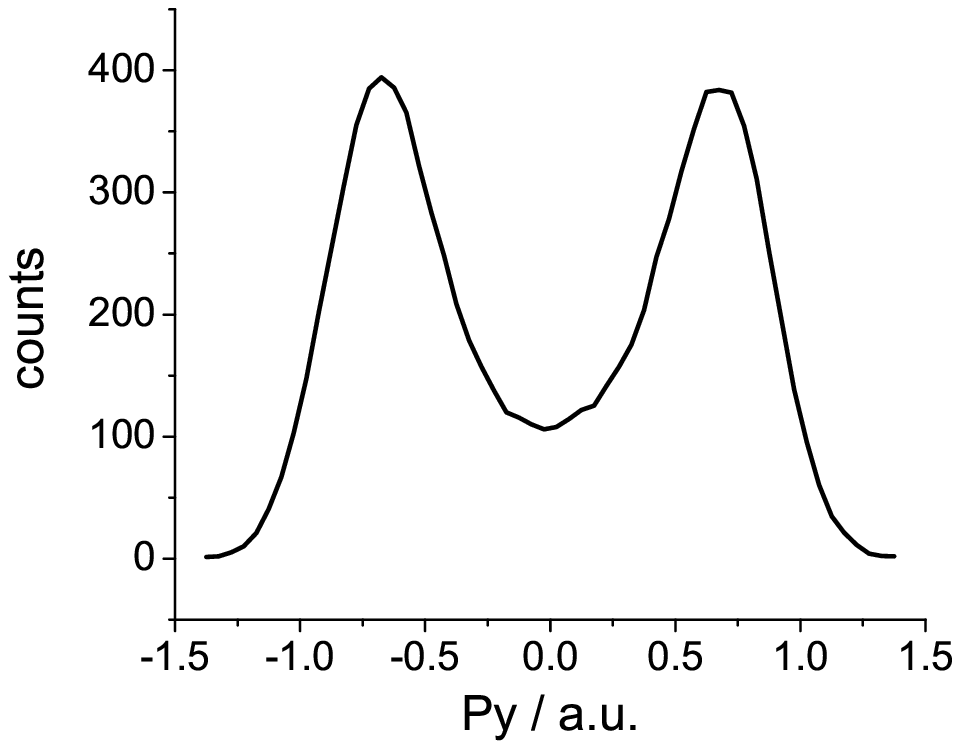}
  \caption{Left: 2D momentum distribution of singly charged ions at 1 PW/cm$^2$. Center: Projection of the 2D momentum distribution on the x direction. Right: Projection of the 2D momentum distribution on the y direction.}\label{f.SI}
\end{figure*}

To test the above Simpleman analytical theory, numerical experiments have been performed using the classical ensemble method. This method has been described in detail elsewhere \cite{Panfili-etal}.

The first step is to generate a microcanonical ensemble of classically modeled atoms \cite{Abrines}. In this paper, a model atom is generated with three active electrons. So far as we know, full-dimensional atoms with three or more active electrons have only been modeled using classical approaches \cite{Ho-Eberly06,Wasson-Koonin}. The ensemble is generated such that the total energy is fixed for each ensemble member (i.e., each model atom). This total energy $E_{tot}$ can be expressed as
\beq
    E_{tot} = \sum_{i=1}^3 \left( \frac{p_i^2}{2} - \frac{3}{\sqrt{r_i^2+a^2}} \right) + \sum_{i<j} \frac{1}{\sqrt{r_{ij}^2+b^2}}, \label{e.Etot}
\eeq
where $p_i$ and $r_i$ are the momentum and position of the $i$th electron, and $r_{ij}$ is the distance between the $i$th and the $j$th. Note that the Coulomb potential has been softened with parameter $a$ (between ion and electrons) and parameter $b$ (between electrons) to stabilize the classically modeled atom \cite{Su-Eberly}. We have set $a$ to be 1.0 a.u., to prevent autoionization, and $b$ to be 0.1 a.u., to avoid numerical singularities. $E_{tot}$ is set by summing the first three ionization potentials, and we take -4.63 a.u., by choosing to model three electrons in neon. Given the total energy, the positions and momenta of the three electrons within an atom are randomly assigned. A projection of the ensemble phase space onto the x$_1$-x$_2$ plane is shown in Fig. \ref{f.ensemble}. Each black dot represents a model atom (an ensemble member). The distribution shows a butterfly shape with lower probabilities in the first and third quadrants than in the second and fourth quadrants, a manifestation of mutual electron repulsion.

Then a laser pulse is turned on and the motion of the electrons is governed by the TDNE
\beq
    \frac{\mathrm{d} \vec{r}}{\mathrm{d} t} = \frac{\partial H}{\partial \vec{p}}; \quad
    \frac{\mathrm{d} \vec{p}}{\mathrm{d} t} = -\frac{\partial H}{\partial \vec{r}}. \label{e.TDNE}
\eeq
The Hamiltonian including the time-dependent laser interaction is
\beq
    H = H(t) = E_{tot} + \sum_{i=1}^3 \left[ x_i E_x(t) + y_i E_y(t) \right].
\eeq

The TDNEs are integrated numerically from the beginning to the end of the pulse. The laser field is given the common experimental wavelength of 780nm ($\omega =$ 0.0584 a.u.) and an ellipticity of 0.5. The pulse has a sine-squared shape with full duration of 10 optical cycles (FWHM = 5 cycles), as shown in Fig. \ref{f.E2T}. The positions and momenta of electrons are recorded at each time step. We have defined ionization as complete when an electron reaches a distance of 6 a.u. from the ion core \cite{Wang-EberlyPRL09} and we have checked that a slight difference in this definition will not affect our discussion here.

At each time step, the ionization criterion is applied to check each electron's status. If at some time step an electron is detected to reach the 6 a.u. shell (note that an electron cannot reach this distance in the absence of the laser field), this time step is labeled as $t_1$ and the laser field strength at this time is labeled as $E_1$ so
\beq
    E_1 = \sqrt{E_x(t_1)^2+E_y(t_1)^2}. \label{e.E1}
\eeq

Due to the field ellipticity, the possibility of recollision can be safely neglected (recall Fig. \ref{f.Simpleman}). If at some later time step, a second electron is detected to reach the 6 a.u. shell, this time step is labeled as $t_2$ and the laser field strength at this time is labeled as $E_2$, which is defined similarly as Eq. (\ref{e.E1}). The same strategy can also be applied to the third ionized electron, and the ionization time and the ionization field will be labeled as $t_3$ and $E_3$. At the end of the pulse, depending on the ionization results, this model atom will be classified into one of the four possible outcome categories: no ionization, single ionization, double ionization, and triple ionization. Each category will then be analyzed separately.

Our numerical experiment can be ``more than" a real lab experiment. The former can get what the latter can, namely, the end-of-pulse ion momentum distributions, and it can also get what the latter cannot, namely, the ionization fields and the ionization times of electrons recorded during the pulse. Recall that the strategy of our Simpleman analytical theory introduced in the previous section is exactly to find this kind of ionization information from experimentally measured ion momentum distributions. Therefore numerical experiments are ideal to test the validity and precision of the analytical theory: We start from the ``experimental results" (the numerical end-of-pulse ion momentum distributions), apply the analytical theory and get the ionization fields and the ionization times, and then compare these values with the actual values (the numerically recorded values).

\section{multi-ionization results}

\begin{figure*}
  \includegraphics[width=4cm]{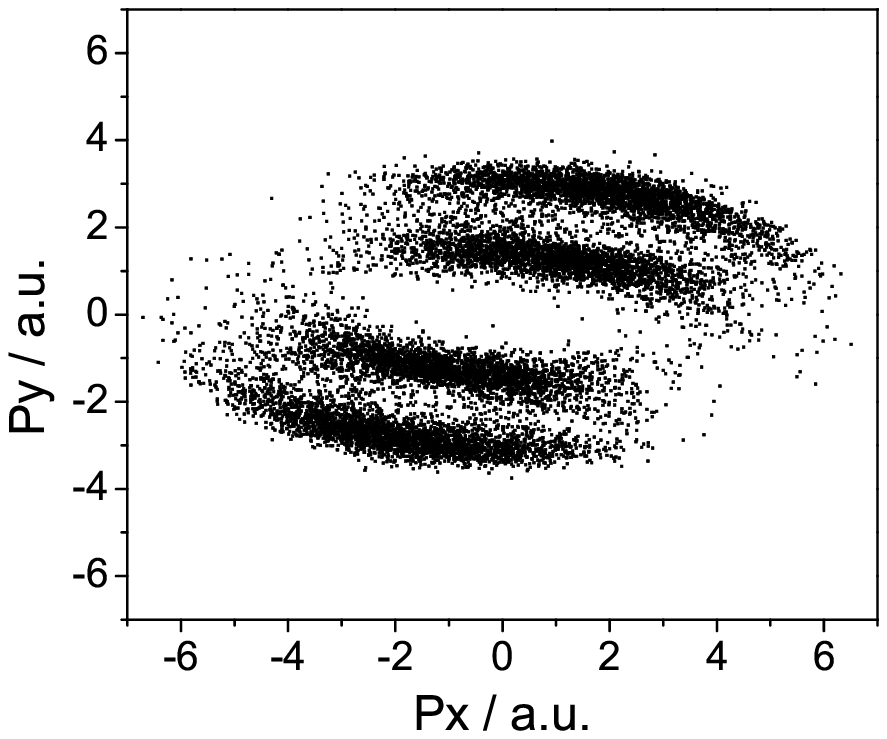}
  \includegraphics[width=4cm]{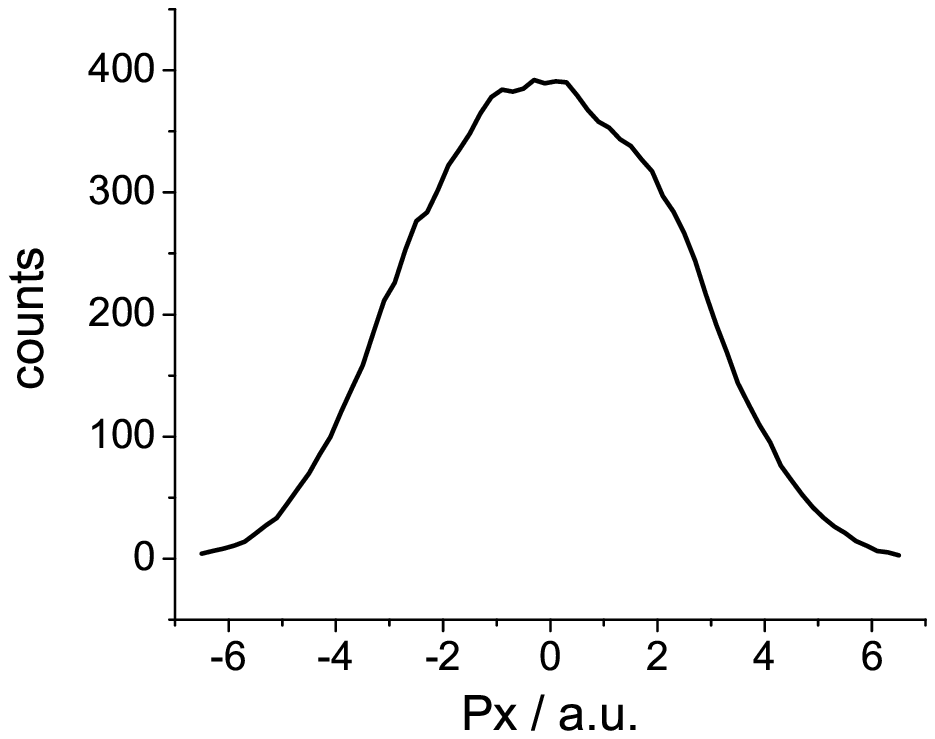}
  \includegraphics[width=4cm]{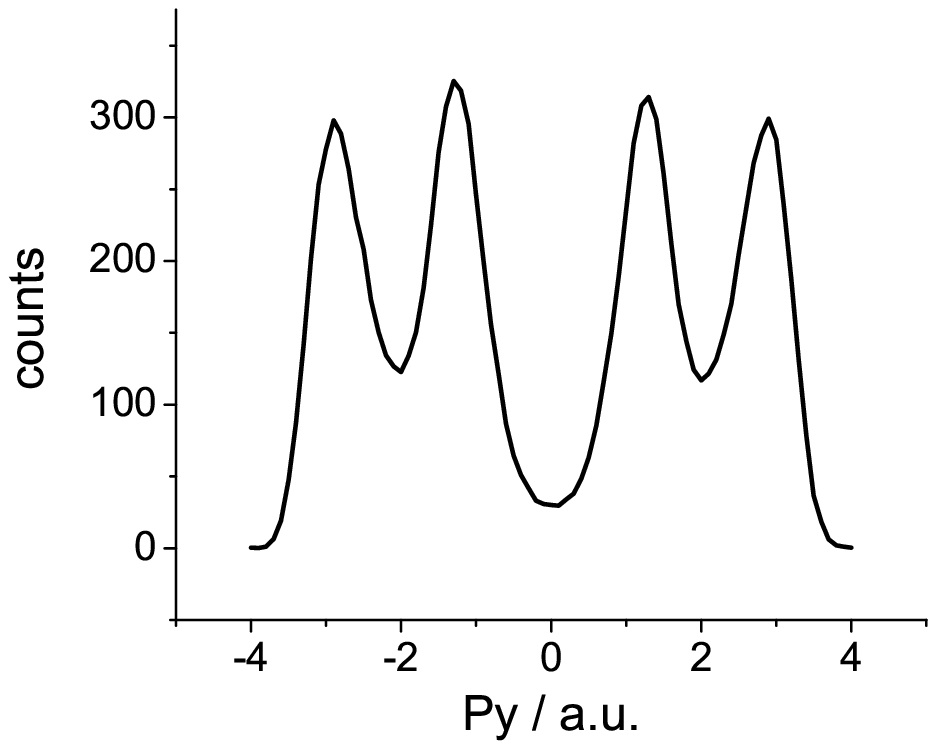}
  \caption{Left: 2D momentum distribution of doubly charged ions at 3 PW/cm$^2$. Center: Projection of the 2D momentum distribution on the x direction. Right: Projection of the 2D momentum distribution on the y direction.}\label{f.DI}
\end{figure*}

To begin, Fig. \ref{f.SI} shows the end-of-pulse momentum distributions of singly charged ions for pulses with peak intensity of 1 PW/cm$^2$. One can find, consistent with the analytical theory, that the momentum distribution along the x direction peaks at zero. The appearance of the shoulders is due to the fact that the 2D ring shape shown in the left panel is not uniformly distributed. Instead, it is denser in the first and the third quadrants than in the second and the fourth quadrants. A clear double peak structure can be seen along the y direction. The peaks are located at $\pm 0.675$ a.u. Using Eq. (\ref{e.E1Peak}) of the Simpleman theory, the ionization field of the electron is expected to be 0.079 a.u. On the other hand, as mentioned above, the classical numerical experiment allows one to know at what time and under what laser field strength an electron was actually ionized. The averaged ionization field of all single ionization events is 0.073 a.u., which deviates less than 10\% from the value inferred from the transverse momentum distribution. The difference may be due to the Coulomb attraction from the ion core, which was taken into account all the time in the numerical experiment but was neglected in the analytical theory. The ionization time found numerically using the method illustrated in Fig. \ref{f.E2T} is 2.76 cycles from the beginning of the pulse, almost exactly the same as the value recorded by the numerical experiment, which is 2.77 cycles from the beginning of the pulse. One can see that under our extension to an elliptically polarized field the Simpleman analytical theory remains valid and precise for electron ionization dynamics.

The single ionization double peak structure has been used by Arissian, et al., to obtain the ionization field \cite{Arissian-etal}. Circular polarization was claimed although in practice, with such high intensity and short pulse duration, elliptical contaminants are usually difficult to avoid. The angular distribution of the 2D momentum distribution has been used by Eckle, et al., to measure the time that an electron needs to tunnel through a Coulomb barrier \cite{Eckle-etal}.

Figure \ref{f.DI} shows the end-of-pulse momentum distribution of doubly charged model neon ions for peak intensity 3 PW/cm$^2$. The 2D momentum distribution shows a four-band structure, corresponding to the four peaks when projecting onto the y axis. The momentum distribution along the x direction shows a broad single peak structure centered at zero. The SDI four-peak structure has been observed in experiment \cite{Maharjan-etal,Pfeiffer-etalNatPh} and explained in detail by our classical ensemble method \cite{Wang-EberlyPRL09}.

Table \ref{t.DI} compares the ionization fields and the ionization times of the two ionized electrons, by locating the peak positions and using the analytical theory, and by records of the numerical experiment. One sees that the analytical theory fits the numerical experiment pretty well. The small discrepancy on $t_2$ between the theory and the numerical experiment is due to the fact that $t_2$ is close to the top of the envelope, which is flat, so the ionization time can have a relatively large uncertainty.

\begin{table}
\begin{tabular}{|c|c|c|}
  \hline
   & \text{ Analytical Theory } & \text{ Numerical Experiment }\\
  \hline
  $E_1$ & 0.093 a.u. & 0.091 a.u. \\
  \hline
  $E_2$ & 0.25 a.u. & 0.22 a.u. \\
  \hline
  $t_1$ & 2.27 cycles & 2.23 cycles \\
  \hline
  $t_2$ & 5.00 cycles & 5.14 cycles \\
  \hline
\end{tabular}
\caption{Comparison of the ionization fields and the ionization times of the two electrons, from the analytical theory and from the numerical experiment.}\label{t.DI}
\end{table}

\begin{figure}[b!]
  \includegraphics[width=8cm]{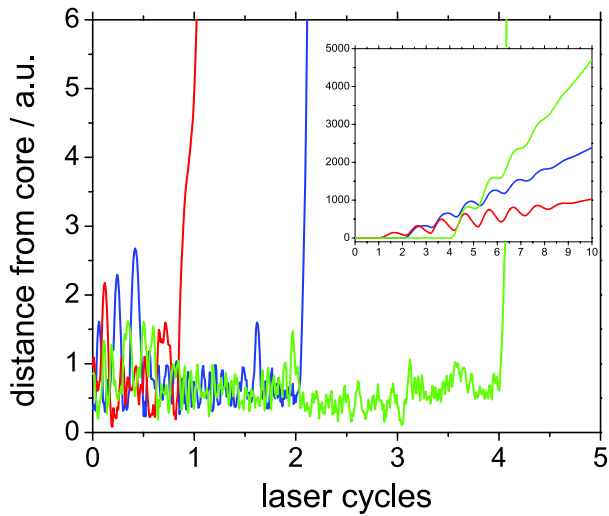}\\
  \caption{A typical triple ionization trajectory. Distances of the three electrons from the ion core are plotted as a function of time, in laser cycles. The inserted figure shows the same trajectory, but in a much larger space scale.}\label{f.triple_traj}
\end{figure}

Recently, Pfeiffer, et al., have used elliptical polarization (with ellipticity value around 0.8) to measure the ionization times of the two electrons in SDI \cite{Pfeiffer-etalNatPh}. The method used therein is different from our method described above. In \cite{Pfeiffer-etalNatPh}, the momenta of the two electrons are measured in addition to the momentum of the ion and the three particles are collected in coincidence, meaning that one must carefully check whether the two electrons and the doubly charged ion are actually from the same atom. A coincidence experiment requires a strict vacuum condition and a very low double ionization rate, such that each time only one atom is ionized, to eliminate possible contaminants from the ionization of neighboring atoms. Even when the experiment is performed with extreme care, false coincidences cannot be fully eliminated \cite{Pfeiffer-etalNJP}. In contrast, we propose in this paper that by just measuring the ion momentum, the same information of the ionization times of the two electrons can be obtained. No electron momenta are needed. No coincidence measurements are needed. Therefore the efficiency of the experiment may be substantially improved and one no longer needs to worry about false coincidences.

\begin{figure}
  \includegraphics[width=4cm]{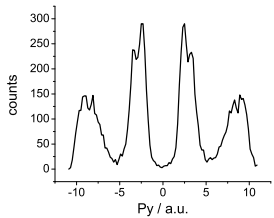}\\
  \caption{Momentum distribution of triply charged ions along the y direction. Eight peaks can be recognized although separations between each two-peak pair is small. Laser peak intensity is 30 PW/cm$^2$.}\label{f.TIPy}
\end{figure}

Finally, we predict similar features in triple ionization. A typical triple ionization trajectory is shown in Fig. \ref{f.triple_traj}. Before emission, electrons are subject to random and fast collisions from the other electrons and the time scale of these internal collisions is much shorter than an optical cycle. After emission, electrons are driven away by the laser field and may travel for a distance on the order of ~1,000 a.u. through the end of a typical pulse. The transverse momentum distribution of triply charged ions generated by 30 PW/cm$^2$ pulses is shown in Fig. \ref{f.TIPy}. Eight peaks can still be recognized although the separations between two neighboring peaks are small. We have also compared the ionization fields and the ionization times of the three ionized electrons obtained from the analytical theory and the numerical experiment, as tabulated in table \ref{t.TI}.

\section{discussion and summary}

In summary, we have focused on elliptical polarization, with which recollisions rarely happen \cite{Wang-EberlyNJP}, although most attention in strong-field atomic physics has been paid to recollision-based physical processes. We show that elliptical polarization has the ability to reveal ionization information that is unreachable with linear polarization. Examples include the ionization fields and the ionization times of emitted electrons.

\begin{table}
\begin{tabular}{|c|c|c|}
  \hline
   & \text{ Analytical Theory } & \text{ Numerical Experiment }\\
  \hline
  $E_1$ & 0.070 a.u. & 0.081 a.u. \\
  \hline
  $E_2$ & 0.34 a.u. & 0.32 a.u. \\
  \hline
  $E_3$ & 0.68 a.u. & 0.68 a.u. \\
  \hline
  $t_1$ & 1.11 cycles & 1.10 cycles \\
  \hline
  $t_2$ & 2.27 cycles & 2.23 cycles \\
  \hline
  $t_3$ & 3.76 cycles & 3.79 cycles \\
  \hline
\end{tabular}
  \caption{Comparison of the ionization fields and the ionization times of the three ionized electrons, from the analytical theory and from the numerical experiment.}\label{t.TI}
\end{table}

We have extended the Simpleman theory to include elliptical polarization and to predict, for the first time, links between the ion momentum distributions and the ionization fields and the ionization times of the emitted electrons. The ion momentum distribution along the minor polarization direction, which is not available with linear polarization, is shown to contain previously unexpected rich ionization information and is predicted to have specific peak structures: a double-peak structure for singly charged ions, a four-peak structure for doubly charged ions, and an eight-peak structure for triply charged ions. The triple ionization eight-peak structure has not yet been observed in experiment.

One should note that the separation between peaks can be enlarged by using a longer wavelength. As predicted by Eq. (\ref{e.SIPy}) and Eq. (\ref{e.DIPy}), the positions of peaks are inversely proportional to $\omega$, thus proportional to the wavelength $\lambda$. Longer wavelengths will be able to resolve close peaks that are not able to be resolved with 800nm, such as the ones for Ar in \cite{Maharjan-etal}. Substantial technical advancements have been made in the direction of longer wavelengths \cite{Agostini-DiMauro} and greater potential in exploring new ionization dynamics is to be expected.

The extension of the Simpleman theory to include elliptical polarization has been tested by our numerical experiments using the classical ensemble method, including three active electrons in the model atom. The numerical experiments confirm the validity and accuracy of the Simpleman analytical theory.

Acknowledgement: This research was supported by DOE Grant No. DE-FG02-05ER15713. We acknowledge helpful discussion and communications with J. Biegert, J. Dura, U. Keller, A. Pfeiffer and D. Villeneuve.

\end{document}